\documentclass[hyper]{JHEP} 
\usepackage{epsfig}
\usepackage{amsmath}
\newcommand\fverb{\setbox\pippobox=\hbox\bgroup\verb}
\newcommand\fverbdo{\egroup\medskip\noindent%
            \fbox{\unhbox\pippobox}\ }

\newcommand\fverbit{\egroup\item[\fbox{\unhbox\pippobox}]}
\newbox\pippobox
\title{D-Brane on Deformed $AdS_3\times S^3$}

\preprint{\hepth{}}

\author{
Malak Khouchen and Josef Kluso\v{n} \\
Institute for Theoretical Physics  and Astrophysics\\
Faculty of Science, Masaryk University\\
Kotl\'{a}\v{r}sk\'{a} 2, 611 37, Brno\\
Czech Republic\\
E-mail: \email{khouchen@mail.muni.cz,klu@physics.muni.cz}}

\abstract{We study D1-brane in $AdS_3\times S^3$ $\kappa$-deformed
background with non-trivial dilaton and Ramond-Ramond fields.  We
consider purely time-dependent and spatially-dependent ansatz where
we study the solutions of the equations of motion for D1-brane in
given background. We find that the behavior of these solutions crucially
depends on the value of the parameter $a$ that was introduced in [arXiv:1411.1066 [hep-th]].}

\def\det{\mathrm{det}}

\def\mL{\mathcal{L}}

\begin{document}

\section{Introduction}

The recent developments in the field of higher-dimensional extended
objects have led to the deep understanding of the superstrings
and supergravity theories. D-branes have been by now  well
understood both from the conformal field theory (CFT) and from the
geometric, target-space viewpoint \footnote{For review and extensive
list of references, see for example
\cite{Johnson:2003gi,Johnson:2000ch}.}. Such hyperplanes are
dynamical rather than rigid and they are defined by the property
that open strings can end on them \cite{Polchinski:1995mt}.
 The incorporation of such D-branes permits
  to argue that the different types of
  string theories are different states of
  a single theory, which also
  contain states with arbitrary configurations of D-branes.
The dynamics of Dp-brane is governed by the action
\begin{equation}
S=S_{DBI}+S_{WZ} \ ,
\end{equation}
where $S_{DBI}$ is Dirac-Born-Infeld action\\
\begin{equation}\label{DBI}
S=-T_{p}\int d^{p+1}\sigma e^{-\Phi} \sqrt{-\det
(g_{\alpha\beta}+b_{\alpha\beta}+2\pi\alpha'F_{\alpha\beta} )} \ ,
 \end{equation}
and $S_{WZ}$ is Wess-Zumino action of the form
\begin{equation}\label{WZ}
S_{WZ}=T_p\int \sum_n C^{(n)} e^{(2\pi\alpha')F+b} \ ,
\end{equation}
where $\sigma^\mu,\mu=0,\dots,p$ label world-volume of Dp-brane,
$\Phi(x)$ is the dilaton and $g_{\alpha\beta},b_{\alpha\beta}$ given
in (\ref{DBI}) are  the pull-backs of the target space metric and
the NS-NS two form field  to the world-volume of Dp-brane
\begin{equation}
g_{\alpha\beta}=g_{MN}\partial_\alpha x^M\partial_\beta x^N \ ,
\quad b_{\alpha\beta}=b_{MN}\partial_\alpha x^M\partial_\beta x^N \
,
\end{equation}
where $x^M(\sigma)$ are embedding coordinates of $D1$-brane. Finally,
$F_{\alpha\beta}=\partial_\alpha A_\beta-\partial_\beta A_\alpha$ is
the field strength for the world-volume gauge field $A_\alpha$. The
coupling of Dp-brane to the Ramond-Ramond fields is expressed
through the Wess-Zumino term (\ref{WZ}) where it is understood that
expressions given there are forms and the multiplications between
them have the form of the wedge product.

Motivated by the recent surge of interest in finding out the
dualities in D-branes and fundamental strings bound states in anti-de
Sitter space and gaining more insight into CFT, we study in this paper the dynamics of $D1$-branes
described by Dirac-Born-Infeld action and Wess-Zumino terms. The
corresponding target-space
geometries are three-dimensional  $\kappa$-deformed $AdS_3\times S^3$ space-time.
Very interesting class of  deformations of target space-time have been introduced in \cite{Delduc:2013qra,Delduc:2014kha}
that preserve the integrability of the two-dimensional quantum field theory on the world
sheet
\footnote{For further works, see 
\cite{Hoare:2014pna,Khouchen:2014kaa,Lunin:2014tsa,Arutyunov:2013ega,Hoare:2015gda,vanTongeren:2015soa,Banerjee:2015nha,Panigrahi:2014sia,Arutyunov:2014jfa,
Hoare:2014oua,Delduc:2014kha,Banerjee:2014bca,Arutyunov:2014cra,Arutynov:2014ota}.}. In the $\kappa$-deformed anti-de Sitter background model, the metric is a direct sum of the
deformed $AdS_n$ and $S^n$ parts and could be truncated from the ten-dimensional metric to $\kappa$-deformed $AdS_3\times S^3$ for example \cite{Hoare:2014pna}. The presence of the deformation parameter $\kappa$ introduces new interesting results that reproduce the ordinary undeformed case in the limit $\kappa\rightarrow 0$ as in  \cite{Khouchen:2014kaa}. Hence, it is interesting to
study the dynamics of  $D1$-brane  in given background as well. In fact,
recently a one-parameter model of the $\kappa$-deformed background $AdS_3\times S^3$ with
non-trivial Ramond-Ramond  (RR) forms and dilaton was proposed in \cite{Lunin:2014tsa}.
A remarkable property of given background is that it depends on parameter $a$ where
it is presumed that $a$ is a particular function of $\kappa$ while
the  full solutions were constructed for the special values
$a=0$ and $a=1$ only.
In our present work, we will use these one-parameter backgrounds to analyze
 static  and time-dependent solutions of $D1$-brane equations of motion
in the background with non-trivial dilaton and
  with RR fields. Our analysis will reveal subtle features.
  Specifically, for the  $\kappa$ deformed $AdS_3$ background with $a=0$
  we will show that $D1$-brane does not see the presence of the singularity of
   the $\kappa$-deformed background and can reach $\rho\rightarrow\infty$ limit. We also find
   that the static solutions are very simple deformations of the static
   solution known as $AdS$ D1-brane in $AdS_3$ background with RR flux.
Since such a solution has not been found in the global coordinates before we
present this result in the Appendix A. Moreover, Appendix B
 exhibits static solutions of  p$D1$-branes in $AdS_3$ space-time with
non-trivial $B_{NS}$  field.
The case of $(1,q)$ string was analyzed previously in \cite{Bachas:2000fr},
but we provide an extended analysis here for the $(p,q)$ string in order to
see the S-duality between given solution and the static D1-brane solution
in undeformed $AdS_3$ space-time with Ramond-Ramond fields.

We also consider static and time-dependent solutions of D1-brane equations
of motion for the $\kappa$-deformed background when the value of the parameter $a$ is
equal to $1$. In this case we find D1-brane cannot cross the singularity
$\rho_c=\frac{L}{\kappa}$ and we also find that the static solutions
is more complicated than in case $a=0$.

The plan of this paper is as follows: In section (\ref{second}) we will consider static $D1$-brane
in the $\kappa$-deformed background  $AdS_3\times S^3$  \cite{Lunin:2014tsa}.
  We study the solutions of the corresponding equations of motions
  and discuss the possibility of the $D1$-brane to reach the $\rho\rightarrow\infty$ limit of the deformed $AdS_3\times S^3$  space.
In section (\ref{third}) we consider  a time-dependent ansatz.
In conclusion (\ref{fourth}) we present summary of our results and their
possible extension.
Finally, some details of the calculations are summarized in the appendices.
In appendix  (\ref{A}) we find static D1-brane solutions in
   non-deformed $AdS_3\times S^3$ background with non-trivial RR fields.
    In appendix  (\ref{B}), we find static solutions of  p$D1$-branes
 in non-deformed $AdS_3$ space-time with non-trivial $B_{NS}$ field
     in global coordinates which is simple generalization of the
     solution found in  \cite{Bachas:2000fr}.
\section{D1-brane in $\kappa$-Deformed Background}\label{second}
In this section, we will study time-independent solutions
of the equations of motion that follow from D-brane actions
(\ref{DBI}) and (\ref{WZ}) in case when $D1$-brane is embedded in
$\kappa$-deformed $AdS_3\times S^3$ background
\cite{Lunin:2014tsa} that has the form
\begin{eqnarray}
ds^2&=&\frac{1}{1-\kappa^2\frac{\rho^2}{L^2}}\left[-\left(1+\frac{\rho^2}{L^2}\right)dt^2+
\frac{d\rho^2}{ 1+\frac{\rho^2}{L^2}}\right]+\rho^2
d\chi^2+\nonumber\\
&+&\frac{1}{1+\kappa^2 \frac{r^2}{L^2}}
\left[\left(1-\frac{r^2}{L^2}\right)d\varphi^2+
\frac{dr^2}{1-\frac{r^2}{L^2}}\right]+r^2d\psi^2 \ , \nonumber \\
\end{eqnarray}
with non-trivial dilaton and Ramond-Ramond fields
\begin{eqnarray}\label{back}
& &a=0 \: \quad
 e^{-2\Phi}=\frac{(1-\kappa^2\frac{\rho^2}{L^2})(1+\kappa^2 \frac{r^2}{L^2})}
{[1-(\frac{\kappa\rho r}{L^2})^2]^2} \
, \nonumber \\
& &C^{(2)}=\frac{1}{L}\frac{1}{1-(\kappa \frac{\rho r}{L^2})^2}
[\rho^2(dt+\kappa L d\varphi)\wedge (d\chi+\kappa \frac{r^2}{L^2}
d\psi)-r^2(L d\varphi-\kappa dt)\wedge ( d\psi+
\kappa \frac{\rho^2}{L^2} d\chi)] \ , \nonumber \\
& &a=1: \quad
e^{-2\Phi}=\frac{(1-\kappa^2\frac{\rho^2}{L^2})(1+\kappa
\frac{r^2}{L^2})} {[1+
\frac{\kappa^2}{L^2}(r^2-\rho^2+r^2\rho^2)]^2} \ , \nonumber \\
&
&C^{(2)}=\frac{\sqrt{1+\kappa^2}}{L(1+\frac{\kappa^2}{L^2}(r^2-\rho^2+\rho^2
r^2))} [\rho^2 dt\wedge d\chi+\kappa[r^2-\rho^2+\frac{1}{L^2}(\rho
r)^2]dt\wedge
d\varphi+\nonumber \\
&+& \frac{\kappa}{L} (\rho r)^2 d\chi\wedge d\psi-r^2L d\varphi
\wedge
d\psi] \ , \nonumber \\
\end{eqnarray}
where $L$ is the inverse curvature scale.
 Consider D1-brane
in given background whose dynamics is governed by the action
\begin{equation}\label{action}
S=-T_{D1}\int d^2\sigma e^{-\Phi} \sqrt{-\det g_{\alpha\beta}
-(2\pi\alpha')^2F_{\tau\sigma}^2 }+T_{D1}\int d^2\sigma C^{(2)}_{MN}\partial_\tau
x^M\partial_\sigma x^N \ .
 \end{equation}
Generally, the equations of motion for $x^M$  that follow from given
action have the form
\begin{eqnarray}\label{eqm}
& &-\partial_\alpha\left[T_{D1} e^{-\Phi}
 \frac{g_{MN}\partial_\beta x^N g^{\beta\alpha}\det g_{\alpha\beta}}
 {\sqrt{-\det g_{\alpha\beta}-(2\pi\alpha')^2F_{\tau\sigma}^2}}
 \right]+T_{D1}
  e^{-\Phi}\partial_M\Phi \sqrt{-\det g_{\alpha\beta}-(2\pi\alpha')^2F_{\tau\sigma}^2}
 +\nonumber \\
 &+&
\frac{T_{D1}}{2} e^{-\Phi}\frac{\partial_M g_{KL}\partial_\alpha x^K\partial_\beta x^L
 g^{\beta\alpha}}{ \sqrt{-\det g_{\alpha\beta}-(2
 \pi\alpha')^2 F_{\tau\sigma}^2}}+
  T_{D1}\partial_M C^{(2)}_{KL}\partial_\tau x^K \partial_\sigma x^L-
 T_{D1}\partial_\alpha[\epsilon^{\alpha\beta}
 C^{(2)}_{MN}\partial_\beta x^N]=0 \ , \nonumber \\
 \end{eqnarray}
 where $\epsilon^{\tau\sigma}=-\epsilon^{\sigma\tau}=1$. On the
 other hand  the equation of motion for $A_\alpha$ implies
\begin{equation}\label{eqA}
T_{D1} \frac{e^{-\Phi}(2\pi\alpha')^2F_{\tau\sigma}}{\sqrt{-\det
g_{\alpha\beta}- (2\pi\alpha')^2F_{\tau\sigma}^2}}=\Pi \ ,
\end{equation}
where $\Pi$ is constant that counts the number of fundamental strings.
 Using this result, we express $F_{\tau\sigma}$ as
\begin{equation}
(2\pi\alpha')F_{\tau\sigma}=\frac{\Pi}{2\pi\alpha'}
\frac{\sqrt{-\det
g_{\alpha\beta}}}{\sqrt{T_{D1}^2e^{-2\Phi}+\frac{\Pi^2}{(2\pi\alpha')^2}}}
\ .
\end{equation}
Then the equations of motion (\ref{eqm}) simplify as
 \begin{eqnarray}\label{eqm1}
& &T_{D1}\partial_\alpha\left[\sqrt{e^{-2\Phi}+\frac{T_{F1}^2}{T_{D1}^2}\Pi^2}
g_{MN}\partial_\beta x^N g^{\beta\alpha}\right]
+T^2_{D1}
  e^{-2\Phi}\partial_M\Phi\frac{ \sqrt{-\det g_{\alpha\beta}}}
 {\sqrt{T_{D1}^2 e^{-2\Phi}+\frac{\Pi^2}{(2\pi\alpha')^2}}}-
 \nonumber \\
 &-&\frac{T_{D1}}{2}\sqrt{e^{-2\Phi}
+\frac{T_{F1}^2}{T_{D1}^2}\Pi^2}
\partial_M g_{KL}\partial_\alpha x^K\partial_\beta x^L
 g^{\beta\alpha}\sqrt{-\det g_{\alpha\beta}}+\nonumber \\
 &+&  T_{D1}\partial_M C^{(2)}_{KL}\partial_\tau x^K \partial_\sigma x^L-
 T_{D1}\partial_\alpha[\epsilon^{\alpha\beta}
 C^{(2)}_{MN}\partial_\beta x^N]=0 \ , \nonumber \\
\end{eqnarray}
where $T_{F1}=\frac{1}{2\pi\alpha'}$.
 This is the form of the equation of motion for D1-brane that we will 
be our starting point.
 If we now return to the specific background given in  (\ref{back}) , we see that since the
background depends on $r$ through its square we find
that the equation of motion for constant $r$ has the
form
\begin{equation}
\frac{\delta \mL}{\delta r^2}r=0
\end{equation}
that has solution $r=0$. In the same way we can show 
that the equations of motions for $\varphi$ and $\psi$ 
have the solutions $\varphi=\psi=0$. In other words we
will not consider solutions with non-trivial behavior on 
deformed $S^{(3)}$. 

\subsection{Static Solutions}
Let us now consider the static $D1$-brane solution when we assume the following
ansatz
\begin{equation}\label{ansstat}
x^0=\tau \ , \quad \chi=\sigma \ , \quad \rho=\rho(\chi) \
\end{equation}
so that
\begin{equation}
g_{\tau\tau}=g_{tt} \ , \quad g_{\sigma\sigma}=
g_{\chi\chi}+g_{\rho\rho}\rho'^2 \ , \quad \rho'\equiv \frac{d\rho}{d\chi} \ .
\end{equation}
For this ansatz the equation of motion for $x^0=t$ is obeyed automatically.
In order to solve for $\rho$ it is more convenient to
consider the equation of motion for $\chi$ since the background fields do not depend on
$\chi$ explicitly. Then, we obtain
\begin{eqnarray}
\sqrt{e^{-2\Phi}+\frac{T_{F1}^2}{T_{D1}^2}\Pi^2}
g_{\chi\chi}\frac{\sqrt{-g_{tt}}}{\sqrt{g_{\chi\chi}+g_{\rho\rho}\rho'^2}}
+
 C^{(2)}_{\chi t}=C \ , C=\mathrm{constant} \ . \nonumber \\
\end{eqnarray}
From given equation we derive the differential equation for $\rho$ in the form
\begin{eqnarray}\label{eqfin}
\rho'^2=\frac{1}{g_{\rho\rho}}
\left[-\frac{(e^{-2\Phi}+\frac{T_{F1}^2}{T_{D1}^2}\Pi^2)g_{tt}g^2_{\chi\chi}}
{(C+C_{t\chi}^{(2)})^2}-g_{\chi\chi}\right]
 \ . \nonumber \\
 \end{eqnarray}
In the following, we will solve this equation for different
background fields by considering two a-families.
\subsubsection{The case a=0}
 Let us begin with the case $a=0$ so that we
have the following background fields
\begin{equation}\label{fieldsa0}
 e^{-\Phi}=\sqrt{1-\frac{\kappa^2}{L^2}\rho^2} \ , \quad
C^{(2)}_{t\chi}=\frac{1}{L}\rho^2 \ .
\end{equation}
Then (\ref{eqfin}) gives
\begin{eqnarray}
\rho'^ 2=\frac{\rho^4(1+\frac{T_{F1}^2}{T_{D1}^2}\Pi^2-
\frac{\kappa^2}{L^2}\rho^2)(1+\frac{\rho^2}{L^2})^2}{C^2(1+\frac{\rho^2}{C
L })^2}- \rho^2 (1-\frac{\kappa^2}{L^2}\rho^2)(1+\frac{\rho^2}{L^2}) \ .
\nonumber \\
\end{eqnarray}
This equation can be integrated at least in principle. However, when
we choose $C=L$ the given equation simplifies considerably
\begin{eqnarray}
\frac{d\rho}{\rho\sqrt{\frac{\rho^2}{L^2}(\frac{T_{F1}^2}{T_{D1}^2}\Pi^2+\kappa^2)-1}}=d\chi
\nonumber \\
\end{eqnarray}
that has a solution\\
\begin{eqnarray}
\rho^2=\frac{L^2}{(\frac{T_{F1}^2}{T_{D1}^2}\Pi^2+\kappa^2)}\frac{1}{\cos^2(\chi-\chi_0)} \ . 
\nonumber \\
\end{eqnarray}
We choose the integration constant by requiring that D1-brane
approaches $\rho\rightarrow \infty$ for
$\chi\rightarrow 0$. Hence, the final result is
\begin{equation}
\rho^2=\frac{L^2}{
(\frac{T_{F1}^2}{T_{D1}^2}\Pi^2+\kappa^2)}\frac{1}{\sin^2\chi } \ . 
\end{equation}
Surprisingly, we find that there is only mirror modification of the
static solution of D1-brane in $AdS_3$ background with non-trivial
 RR fields that is presented in appendix (\ref{A}),
 where this modification is given by the presence
of the deformation parameter $\kappa$. Further, we also see that
D1-brane can be stretched through the singularity
$\rho_c=\frac{L}{\kappa}$ and can reach $\rho\rightarrow \infty$. This is very remarkable result
especially in the light of the solution that we find in case $a=1$.
\subsubsection{The case a=1}
In this case, the relevant
components of the background fields at $r=0$ are
\begin{eqnarray}\label{fieldsa1}
e^{-2\Phi}=\frac{1}{1-\kappa^2\frac{\rho^2}{L^2}} , \quad
C^{(2)}_{t\chi}=
\frac{\sqrt{1+\kappa^2}}{1-\kappa^2\frac{\rho^2}{L^2}}\frac{\rho^2}{L}
\ .
\nonumber \\
\end{eqnarray}
Again (\ref{eqfin}) gives\\
 \begin{eqnarray}
 \rho'^ 2=
 \frac{(1+\frac{\rho^2}{L^2})^2 \rho^4
 (\frac{1}{1-\kappa^2\frac{\rho^2}{L^2}}+\frac{T_{F1}^2}{T_{D1}^2}\Pi^2)}{
 (C+\frac{\sqrt{1+\kappa^2}}{1-\kappa^2\frac{\rho^2}{L^2}}\frac{\rho^2}{L})^2}
 -\rho^2(1-\kappa^2\frac{\rho^2}{L^2})(1+\frac{\rho^2}{L^2})
\nonumber \\
\end{eqnarray}
We simplify this equation by choosing
$C=\frac{L}{\sqrt{1+\kappa^2}}$. Hence, the previous equation has
the form
\begin{eqnarray}
\rho^2_{max}\rho_{min}\frac{d\rho}{\rho(\rho^2_{max}-\rho^2)\sqrt{\rho^2-\rho^2_{min}}}=
d\chi \ ,
\nonumber \\
\end{eqnarray}
where
\begin{equation}
\rho^2_{max}=\frac{L^2}{\kappa^2} \  ,   \rho^2_{min}=\frac{L^2}{(1+\kappa^2)
\Pi^2 \frac{T_{F1}^2}{T_{D1}^2}} \ .
\end{equation}
Solving this equation, we get\\
\begin{eqnarray}\label{sola1}
\tan^{-1}\frac{\sqrt{\rho^2-\rho^2_{min}}}{\rho_{max}}+
\frac{\rho_{min}}{2\sqrt{\rho^2_{max}-\rho^2_{min}}}\ln
\left[\frac{\sqrt{\rho^2_{max}-\rho^2_{min}}+\sqrt{\rho^2-\rho^2_{min}}}
{\sqrt{\rho^2_{max}-\rho^2_{min}}-\sqrt{\rho^2-\rho^2_{min}}}\right]=\chi-\chi_0
\ .
\nonumber \\
\end{eqnarray}
We choose the integration constant $\chi_0$ in such a way
that for $\rho\rightarrow \rho_{min}$ , $\chi\rightarrow \frac{\pi}{2}$. Then we obtain
\begin{equation}
 \chi_0=\frac{\pi}{2} \ .
\end{equation}
From (\ref{sola1}), we see that $D1$-brane does not reach the maximum
value $\rho_{max}$ for $\chi$ in the interval $\chi\in (0,2\pi)$.
More precisely, $D1$-brane reaches the maximum value at $\rho_{max}$
for $\chi\rightarrow -\infty$ that implies that $D1$-brane has to wrap
compact $\chi$ direction infinitely many times. We also see that now
$D1$-brane does not cross the singularity at $\rho_c=\frac{L}{\kappa}$.
In summary, we see qualitative different behaviors of these two
solutions corresponding to the cases $a=0$ and $a=1$. We will also
see this difference in case of pure time-dependent solutions that
will be analyzed in the next section.
\section{Time-Dependent Solution}\label{third}
In order to find time-dependent solution, we consider an ansatz
\begin{equation}
t=\tau \ , \chi=\sigma \ , \rho=\rho(t)
\end{equation}
so that
\begin{equation}
g_{\tau\tau}=g_{tt}+g_{\rho\rho}(\dot{\rho})^2 \ , \quad
g_{\sigma\sigma}=g_{\chi\chi} \ .
\end{equation}
For such ansatz, we find that the equation of motion
for $\chi$ is automatically obeyed. On the other hand, the equation
of motion for $t$ gives
\begin{eqnarray}
\sqrt{e^{-2\Phi}+\frac{T_{F1}^2}{T_{D1}^2}\Pi^2}
\frac{g_{tt}\sqrt{g_{\chi\chi}}}{\sqrt{-g_{tt}-g_{\rho\rho}(\dot{\rho})^2}}-
C_{t\chi}^{(2)}=C \ , C=\mathrm{const} \
\nonumber \\
\end{eqnarray}
that implies following  differential equation for $\dot{\rho}$
\begin{eqnarray}\label{rhodot}
\dot{\rho}=\frac{\sqrt{-g_{tt}}}{\sqrt{g_{\rho\rho}}}
\sqrt{1+\frac{g_{tt}g_{\chi\chi}
(e^{-2\Phi}+\frac{T_{F1}^2}{T_{D1}^2}\Pi^2)}{(C+C_{t\chi}^{(2)})^2}}
\nonumber \ . 
\\
\end{eqnarray}
In the following we will consider  two different one-parameter a-families of background
fields.\\
\subsection{The case a=0}
Substituting the fields of (\ref{fieldsa0}) in (\ref{rhodot}), we obtain
\begin{eqnarray}\label{rhodot0}
\dot{\rho}=(1+\frac{\rho^2}{L^2}) \sqrt{
1-\frac{\rho^2(1+\frac{\rho^2}{L^2})(1+\frac{T_{F1}^2}{T_{D1}^2}\Pi^2-\kappa^2
\frac{\rho^2}{L^2})}
{(1-\kappa^2\frac{\rho^2}{L^2})(C+\frac{\rho^2}{L})^2}} \nonumber
\\
\end{eqnarray}
Let us impose the condition $C=L$
that simplifies the given equation considerably.
 For this condition, the turning point at which $\dot{\rho}=0$ will be at
\begin{eqnarray}
 \rho_{max}=\frac{L}{\sqrt{\Pi^2\frac{T_{F1}}{T_{D1}}+\kappa^2}}\nonumber \\
\end{eqnarray}
which is less than $\frac{L}{\kappa}$. Hence, we see that in this
case the $D1$-brane does not cross the singularity at
$\rho_c=\frac{L}{\kappa}$.

On the other hand let us consider  the case when $\Pi=0$. From
(\ref{rhodot0}), we obtain
\begin{eqnarray}
\dot{\rho}=(1+\frac{\rho^2}{L^2}) \sqrt{
1-\frac{\rho^2(1+\frac{\rho^2}{L^2})} {(C+\frac{\rho^2}{L})^2}}
\nonumber
\\
\end{eqnarray}
It seems interesting  that now  the expressions containing
the deformation parameter $\kappa$ disappear. The turning point is
at
\begin{eqnarray}
\rho^2_{t.p.}=\frac{C^2}{1-2\frac{C}{L}} \
\end{eqnarray}
this implies that in order to have real solution we have to demand
that $C<\frac{L}{2}$. For $C=\frac{L}{2}$, we realize that the
$D1$-brane reaches  $\rho\rightarrow\infty$ asymptotically. More explicitly, in such case we can
easily integrate the differential equation with the result
\begin{equation}
2\rho-L\tan^{-1}\frac{\rho}{L}=t \ ,
\end{equation}
and we see that for $t\rightarrow \infty$ $D1$-brane approaches ($\rho=\infty$).\\
 Finally,  for $C>\frac{L}{2}$ the $D1$-brane
reaches  $\rho\rightarrow\infty$ since the expression 
under the square root is then always positive
without any restrictions on the radial coordinate $\rho$.
In other words, $D1$-brane with zero electric field can probe the space-time
beyond the singularity $\rho_c=\frac{L}{\kappa}$ as well.
\\

\subsection{The case a=1}
In this case, substituting the fields of (\ref{fieldsa1}) in (\ref{rhodot}),
 the differential equation has the form
\begin{eqnarray}
\dot{\rho}=(1+\frac{\rho^2}{L^2})
\sqrt{1-\frac{(1+\frac{\rho^2}{L^2})\rho^2
(1+\frac{T_{F1}^2}{T_{D1}^2}\Pi^2(1-\kappa^2\frac{\rho^2}{L^2}))}
{(C(1-\kappa^2\frac{\rho^2}{L^2})+
\sqrt{1+\kappa^2}\frac{\rho^2}{L})^2}} \nonumber \\
\end{eqnarray}
We are interested in the special case when $\Pi=0$. In this case, we
find the turning point at
\begin{eqnarray}
\rho^2_{t.p.}= \frac{-(2CA-1)\pm
\sqrt{(2CA-1)^2-4C^2(A^2-\frac{1}{L^2})}}{2(A^2-\frac{1}{L^2})}
\nonumber \\
\end{eqnarray}

where $A=\frac{\sqrt{1+\kappa^2}}{L}-C\frac{\kappa^2}{L^2}$.
After the analysis of the expression under the square
root (the discriminant), we realize that it is always positive.
Then we have to consider two cases.

In the first case, $A^2-\frac{1}{L^2}<0$ then to have an overall positive quantity, we
require a condition
\begin{equation}
C\in (L (\frac{\sqrt{1+\kappa^2}-1}{\kappa^2}), L
(\frac{\sqrt{1+\kappa^2}+1}{\kappa^2}))
\end{equation}
In the second case we have $A^2-\frac{1}{L^2}>0$ that gives
\begin{equation}
C\in (-\infty, L( \frac{\sqrt{1+\kappa^2}-1}{\kappa^2})\cup
(L(\frac{\sqrt{1+\kappa^2}+1}{\kappa^2},\infty)
\end{equation}
In this case however we have also to demand that
\begin{equation}
2CA-1<0
\end{equation}
that implies
\begin{eqnarray}
C\in [(-\infty,
\frac{L}{2\kappa^2}(\sqrt{1+\kappa^2}-\sqrt{1-\kappa^2})\cup
(\frac{L}{2\kappa^2}(\sqrt{1+\kappa^2}+\sqrt{1-\kappa^2}),\infty))]
\nonumber \\
\end{eqnarray}
Now, however we find that 
the second
condition is always obeyed since the second interval is included in the first. Therefore, there always exist
real roots corresponding to the turning points $\rho_{r.t.}$.

Let us try to determine the value of the turning point for large
$\frac{C}{L}\gg 1$ (large energy limit). In this case, we can write $A\approx
-\frac{C}{L^2}\kappa^2$ and we obtain
\begin{eqnarray}
\rho^2_{r.t.}=\frac{L^2}{\kappa^2}\left(1+\frac{L^2}{2C^2\kappa^4}\pm
\frac{L}{C}\frac{\sqrt{1+\kappa^2}}{2\kappa^2}\right) \nonumber \\
\end{eqnarray}
so that when we restrict to the terms linear in $\frac{L}{C}\ll 1$
we obtain two roots
\begin{equation}
\rho^2_{max}=\frac{L^2}{\kappa^2}\left(1+\frac{L}{C}\frac{\sqrt{1+\kappa^2}}{2\kappa^2}\right)
\ , \nonumber \\
\rho^2_{min}=\frac{L^2}{\kappa^2}\left(1-\frac{L}{C}\frac{\sqrt{1+\kappa^2}}{2\kappa^2}\right) \ .
\end{equation}
In other words we find two situations. In the first case D1-brane is
in the region below the singularity $\rho^2_c=\frac{L^2}{\kappa^2}$ and
can reach its turning point at $\rho^2_{min}$ and then it returns
back. In the second case, D1-brane is in the region
$\rho^2>\rho^2_{max}$ i.e. beyond the singularity. However, it is important that in both of these
cases, D1-brane cannot go through the singularity.

Finally, we compare this result with the analysis of the time-dependent solution of the fundamental string in $\kappa$-deformed
background. Recall that the fundamental string is described by the
Nambu-Gotto action
\begin{equation}
S=-T_{F1}\int d\tau d\sigma \sqrt{-\det g_{\alpha\beta}} \ ,
g_{\alpha\beta}=g_{MN}\partial_\alpha x^M\partial_\beta x^N \ .
\end{equation}
The equation of motion for $t$ for the time dependent ansatz again
implies
\begin{eqnarray}
g_{tt}g^{\tau\tau}\sqrt{-g_{\chi\chi}(g_{tt}+g_{\rho\rho}\dot{\rho}^2)}=C \ , C=\mathrm{constant} \ .
\nonumber \\
\end{eqnarray}
Solving given equation for $\dot{\rho}$ we obtain
\begin{eqnarray}
\dot{\rho}=\frac{1}{CL}(1+\frac{\rho^2}{L^2})
\sqrt{\frac{(\rho^2_--\rho^2)(\rho^2+\rho^2_+)}{1-\kappa^2\frac{\rho^2}{L^2}}}
\nonumber \\
\end{eqnarray}
where
\begin{eqnarray}
\rho^2_-=\frac{L^2}{2}\left[-(1+\kappa^2\frac{C^2}{L^2})+
\sqrt{(1+\kappa^2\frac{C^2}{L^2})^2+4\frac{C^2}{L^2}}\right] \ ,
\nonumber
\\
\rho^2_+=\frac{L^2}{2}\left[(1+\kappa^2\frac{C^2}{L^2})+
\sqrt{(1+\kappa^2\frac{C^2}{L^2})^2+4\frac{C^2}{L^2}}\right] \ .
\nonumber
\end{eqnarray}
Note that for large $C$, $\rho_-$ has the form
\begin{eqnarray}
\rho^2_-=L^2(\frac{1}{\kappa^2}-\frac{L^2}{C^2\kappa^6}) \nonumber
\\
\end{eqnarray}
We see that the allowed regions for the propagation of the string is
$(0,\rho^2_-)$ and $(\frac{L^2}{\kappa^2},\infty)$. In other words,
strings cannot cross the singularity at $\rho_c^2=\frac{L^2}{\kappa^2}$ when
it is originally confined in the region around the point $\rho=0$.
On the other hand, for $\frac{C}{L}\ll 1$ we obtain $\rho^2_-\approx
C^2 \ll L^2$ and the string is confined in the region around
$\rho=0$ there is no sign of the deformation of the target space-time. 
\section{Conclusion}\label{fourth}
In this paper we have studied the dynamics of $D1$-brane
in $\kappa$-deformed  $AdS_3\times S^3$
background with non-trivial dilaton and Ramond-Ramond fields
\cite{Lunin:2014tsa}.
We have found that the background
   with $a=0$ possesses many interesting properties.
We have shown  that the static solution of  $D1$-brane in
  the presence of RR-charges can reach  $\rho\rightarrow\infty$ limit of
the deformed    $AdS_3\times S^3$ space-time  and that given
solution is a slight modification from the AdS $D1$-brane solution
in undeformed $AdS_3\times S^3$ background with Ramond-Ramond flux
that is found in Appendix (\ref{A}). Moreover, it is also very  interesting
that the time dependent solution does not see the presence of the
singularity at $\rho_c=\frac{L}{\kappa}$. In other words, $D1$-brane in deformed
$AdS_3\times S^3$ space-time
can cross given singularity and reach  $\rho\rightarrow\infty$. Hence, $D1$-brane can be considered as
natural probe of given space-time.

These results are in sharp contrast with the case $a=1$ where we
have shown that the $D1$-brane does not reach the singularity.
Explicitly, it was shown that in such conditions the $D1$-brane can
move in the region beyond the singularity or in  a region below the
singularity, but it can not cross the singularity in both situations. The
latter result was confirmed by analyzing the dynamics of the
fundamental string in given background. Again, it was demonstrated
that a string originally confined in the region around $\rho=0$ can
not cross the singularity.

 We have further examined the static gauge ansatz of
$pD1$-branes bound to $q$ fundamental strings with non-trivial NS-NS flux 
 in global coordinates. After solving the equations of
motion, we were able to generalize Bachas result
\cite{Bachas:2000fr} for the constant $C$ that determines the radius
of AdS. We have shown that $C$ is proportional to the number of
fundamental strings in the bound state and inversely proportional to
the number of $D1$-branes. Finally, we considered the time-dependent
solution of $pD1$-branes bound to $q$ fundamental strings in the same
background. We were able to show that in the limit
$T_{(p,q)}\rightarrow qT_F$ 
 the fundamental string can indeed reach $\rho\rightarrow\infty$.

The present analysis can be
 extended in various directions. First of
all it would be very interesting and challenging to study the
dynamics of $pD1$-branes in the $\kappa$-deformed $AdS_3\times S^3$
with complex deformation parameter. Further one can try to find the
complete solution with arbitrary parameter $a$ then proceed with a
similar analysis as we did in this paper. It would be also interesting
to perform analysis of D1-brane and fundamental string configurations
that could describe Wilson loops in dual field theory. We hope to return
to these problems in future.

\vskip .5in \noindent {\bf Acknowledgement:}

 This work   was
supported by the Grant agency of the Czech republic under the grant
P201/12/G028. \vskip 5mm

\begin{appendix}
\section{D1-brane as Probe of $AdS_3\times S^3$ Background with Ramond-Ramond Background}
\label{A}
In this appendix we consider the static solution of
D1-brane equations of motions in the  non deformed
$AdS_3\times S^3$ background with non-zero Ramond Ramond field.
Let us be more explicit and consider the case of the near horizon limit
of D1-D5 brane system that in global coordinates has the form
\footnote{We follow the conventions used in
\cite{Raeymaekers:2006np,Janssen:2004jz}. }
\begin{eqnarray}
ds^2&=&-(1+\frac{\rho^2}{L^2})dt^2+(1+\frac{\rho^2}{L^2})^{-1}d\rho^2+
\rho^2 d\varphi^2+\nonumber \\
&+&L^2(d\theta^2+\cos^2\theta d\phi^2+\sin^2\theta
d\chi^2) \ ,
\nonumber \\
e^{\Phi}&=&\mathcal{R}^2 \ , \quad
C_{t\varphi}^{(2)}=-\frac{Q_5}{L^3}\rho^2 \ , \quad
C_{\phi\chi}^{(2)}=Q_5\sin^2\theta \ ,
\nonumber \\
\end{eqnarray}
where $\varphi,\phi,\chi,y_m\in[0,2\pi] $ and $\theta\in [0,\pi]$
and where
\begin{equation}
\mathcal{R}^2=\sqrt{\frac{Q_1}{Q_5}} \ , \quad L^2=\sqrt{Q_1 Q_5} \
.
\end{equation}
The equations of motion
for $D1$-brane in given background have the form
%
%

\begin{eqnarray}\label{eqXM}
& &\frac{T_{D1}}{2}\frac{e^{-\Phi}\partial_M g_{KL}\partial_ \alpha
x^K\partial_\beta x^L g^{\beta\alpha}\det g}{\sqrt{-\det
g_{\alpha\beta}-(2\pi\alpha')^2 F_{\tau\sigma}^2}}
-\partial_\alpha\left[\frac{T_{D1}e^{-\Phi} g_{MN}\partial_\beta
x^N g^{\beta\alpha}\det g}{\sqrt{-\det g-(2\pi\alpha')^2
F_{\tau\sigma}^2}}\right]-\nonumber \\
&-&\partial_\alpha [C_{MN}\partial_\beta
x^N\epsilon^{\alpha\beta}]=0 \ , \nonumber \\
\end{eqnarray}
while the equation of motion for $A_\alpha$  again implies
\begin{eqnarray}
T_{D1}\frac{e^{-\Phi}(2\pi\alpha')^2 F_{\tau\sigma}} {\sqrt{-\det
g_{\alpha\beta}-(2\pi\alpha')^2(F_{\tau\sigma})^2}}=\Pi \nonumber \\
\end{eqnarray}
Let us now presume an ansatz where the $D1$-brane is wrapping $\tau$
and $\varphi$ directions and where $\rho=\rho(\sigma)$,
\begin{equation}\label{ansRR}
x^0=\tau \ , \varphi=\sigma \ , \rho=\rho(\varphi)
\end{equation}
where  we use the
notation $\sigma^0=\tau, \sigma^1=\sigma$ keeping in mind that
$\sigma$ is dimensionless.
 For the ansatz (\ref{ansRR}), the equation
of motion (\ref{eqXM}) for $M=0$ is automatically satisfied while that of $\varphi$ implies
\begin{equation}
\sqrt{T_{D1}^2+\left(\frac{\Pi}{2\pi\alpha'}\right)^2}e^{-\Phi}g_{\varphi\varphi}
\sqrt{-g}g^{\sigma\sigma}+T_{D1}\frac{Q_5}{L^3}\rho^2=C \ ,
C=\mathrm{constant} \
\end{equation}
that can be solved for $\rho'$ as
\begin{eqnarray}
\rho'^2=\frac{\rho^4(1+\frac{\rho^2}{L^2})^2} {C'^
2(1-\frac{K}{LC'}\rho^2)^2}-
\rho^2(1+\frac{\rho^2}{L^2}) \  , \nonumber \\
\nonumber \\
\end{eqnarray}
where $K^2=\frac{T_{D1}^2}{(T_{D1}^2+(\frac{\pi}{2\pi\alpha'})^2)}$
 and where now $\rho'\equiv \frac{d\rho}{d\varphi}$.
Let us now choose the constant $C'$ in such a way that
\begin{equation}
C'=-KL
\end{equation}
when the equation above simplifies considerably
\begin{equation}
\rho'=\frac{\rho}{KL}\sqrt{\rho^2(1-K^2)-K^2L^2 }
\end{equation}
and hence we find the solution
\begin{equation}
\frac{\rho^2}{L^2}=\frac{K^2 L^2}{1-K^2L^2}\frac{1}{
\cos^2(\varphi-\varphi_0)} \ .
\end{equation}
We again choose the integration constant that for
$\varphi\rightarrow 0$, the system approaches
$\rho\rightarrow \infty$ so that
\begin{equation}\label{AdS2RR}
\frac{\rho^2}{L^2}=
\frac{1}{\Pi^2}\frac{T_{D1}^2}{T_{F1}^2}\frac{1}{\sin^2\varphi} \ .
\end{equation}
This is the solution corresponding to AdS D1-brane in
$AdS_3$ background with non-trivial RR fields. In the next appendix
we show that given configuration is S-dual to the specific
bound state of D1-branes and fundamental strings in $AdS_3$ background
with non-trivial $B_{NS}$ two form.

%
\section{$p$D1-branes in $AdS_3$ with $B_{NS}$ field in
global coordinates}\label{B}
 Let us now consider a collection of $p$D1-branes in $AdS_3$ background with the  metric
\begin{equation}\label{AdS3}
ds^2=-(1+\frac{\rho^2}{L^2})dt^2+\frac{1}{1+\frac{\rho^2}{L^2}}d\rho^2+
\rho^2 d\chi^2
\end{equation}
 but with
non-zero $B_{NS}$ field in the form
\begin{equation}\label{BNS}
B=\frac{\rho^2}{L}  d\chi \wedge dt \ .
\end{equation}
Since we are interested in the collective dynamics
of the bound state of $p$ D1-branes it is clear that given action
is the standard DBI action multiplied with the number
$p$ so that
\begin{equation}
S=-pT_{D1}\int d\tau d\sigma \sqrt{-\det
g_{\alpha\beta}-(b_{\tau\sigma}+2\pi\alpha'F_{\tau\sigma})^2} \ .
\end{equation}
Note that the equations of motion for $x^M$ that follow from given
action have the form
\begin{eqnarray}
& &-\partial_\alpha\left[p T_{D1}\frac{g_{MN}\partial_\beta x^N
g^{\beta\alpha}\det g_{\alpha\beta}}{ \sqrt{-\det g_{\alpha\beta}
-(b_{\tau\sigma}+2\pi\alpha'F_{\tau\sigma})^2}}\right] +\frac{p
T_{D1}}{2} \frac{\partial_M g_{KL}\partial_\alpha x^K \partial_\beta
x^L g^{\beta\alpha}\det g_{\alpha\beta}} { \sqrt{-\det
g_{\alpha\beta} -(b_{\tau\sigma}+2\pi\alpha'F_{\tau\sigma})^2}}-
\nonumber \\
& &-pT_{D1}\partial_\tau
\left[\frac{(b_{\tau\sigma}+(2\pi\alpha')F_{\tau\sigma})b_{MN}\partial_\sigma
x^N}{ \sqrt{-\det g_{\alpha\beta}
-(b_{\tau\sigma}+2\pi\alpha'F_{\tau\sigma})^2}}\right]+
pT_{D1}\partial_\sigma
\left[\frac{(b_{\tau\sigma}+(2\pi\alpha')F_{\tau\sigma})b_{MN}\partial_\tau
x^N}{ \sqrt{-\det g_{\alpha\beta}
-(b_{\tau\sigma}+2\pi\alpha'F_{\tau\sigma})^2}}\right]+
\nonumber \\
& &+p
T_{D1}\frac{(b_{\tau\sigma}+(2\pi\alpha')F_{\tau\sigma})\partial_M
b_{KL}\partial_\tau x^K \partial_\sigma x^L} { \sqrt{-\det
g_{\alpha\beta} -(b_{\tau\sigma}+2\pi\alpha'F_{\tau\sigma})^2}}=0
\nonumber \\
\end{eqnarray}
 The equation of motion for $A_\alpha$ implies
\begin{equation}
pT_{D1}\frac{b_{\tau\sigma}+2\pi\alpha' F_{\tau\sigma}}{ \sqrt{-\det
g_{\alpha\beta}-(b_{\tau\sigma}+2\pi\alpha'F_{\tau\sigma})^2}}
=\frac{q}{2\pi\alpha'} \ ,
\end{equation}
where $q$ is the number of fundamental strings bound to $p$
$D1$-branes. Note that using the previous result we can express
$b_{\tau\sigma}+(2\pi\alpha') F_{\tau\sigma}$ as
\begin{equation}
b_{\tau\sigma}+(2\pi\alpha')F_{\tau\sigma}= \frac{qT_{F1}\sqrt{-\det
g_{\alpha\beta}}}{\sqrt{p^2 T_{D1}^2+q^2 T_{F1}^2}}
\end{equation}
 Let us
now choose the following ansatz
\begin{equation}
t=\tau , \quad  \chi= \sigma \ , \rho=\rho(\sigma) \ .
\end{equation}
In this case we find that the equation of motion for $t$ is
automatically obeyed while the equation of motion for $\chi$ implies
\begin{eqnarray}\label{ptqf}
\sqrt{(pT_{D1})^2+\left(\frac{q}{2\pi\alpha'}\right)^2}g_{\chi\chi}g^{\sigma\sigma}\sqrt{-\det
g} +\frac{q}{2\pi\alpha'} b_{\varphi
t}=\sqrt{(pT_{D1})^2+\left(\frac{q}{2\pi\alpha'}\right)^2}C \ ,
\nonumber \\
\end{eqnarray}
where $C$ is a constant. From given equation we obtain differential
equation for $\rho'$
\begin{equation}
\rho'^2=\frac{\rho^4(1+\frac{\rho^2}{L^2})^2}{(C-\frac{qT_{F1}}{
\sqrt{p^2T_{D1}^2+q^2
T_{F1}^2}L}\rho^2)^2}-\rho^2(1+\frac{\rho^2}{L^2}) \ .
\end{equation}
If we choose the integration constant $C$ to be equal to
\begin{equation}
C=-\frac{L q T_{F1}}{\sqrt{p^2 T_{D1}^2+q^2 T_{F1}^2}} \
\end{equation}
we find simple differential equation for $\rho $
\begin{equation}
\rho'=\rho\sqrt{\frac{\rho^2}{L^2}\frac{p^2 T_{D1}^2}{q^2
T^2_{F1}}-1}
\end{equation}
that has  solution
\begin{equation}
\frac{\rho}{L}=\frac{q T_{F1}}{pT_{D1} }\frac{1}{\sin\chi} \
\end{equation}
which is  the generalization of the solution found in
\cite{Bachas:2000fr} to the case of the bound state of $p$ D1-branes
and $q$ fundamental strings. Note that the special case when
we have $p=\Pi$ D1-branes and $q=1$ fundamental strings is S-dual
to the solution found in the previous section which is the bound
state of single D1-brane and $\Pi$ fundamental strings.
It is also instructive to consider time-dependent solution
corresponding to the motion of the bound state of $p$D1-branes and
$q$ fundamental strings in given background when we consider an
ansatz
\begin{equation}
x^0=\tau \ , \quad \rho=\rho(\tau)\ , \chi=\sigma \ .
\end{equation}
Then the equation of motion for $t$ implies
\begin{equation}\label{eqbnst}
\sqrt{p^2 T_{D1}^2+q^2 T_{F1}^2}g_{tt}g^{\tau\tau}\sqrt{-\det g_{\alpha\beta}}
-qT_{F1}b_{t\chi}=C
\end{equation}
and hence we obtain
\begin{eqnarray}\label{dottfs}
\dot{\rho}=\frac{\sqrt{-g_{tt}}}{\sqrt{g_{\rho\rho}}}
\sqrt{1+\frac{(p^2 T_{D1}^2+q^2 T_{F1}^2)g_{tt}g_{\chi\chi}} {(C-q
T_{F1} b_{t\chi})^2}} \ . 
\end{eqnarray}
For the background given in (\ref{AdS3}) and (\ref{BNS}) we obtain
that there is a turning point at
\begin{eqnarray}
\rho^2=\frac{-( p^2 T_{D1}^2+q^2 T_{F1}^2-2 \frac{C}{L}q T_{F1})+\sqrt{(
p^2 T_{D1}^2+q^2 T_{F1}^2-2 \frac{C}{L}q T_{F1})^2+4p^2
\frac{C^2}{L^2}T_{D1}^2}}{2\frac{p^2 T_{D1}^2}{L^2}} \ .  \nonumber
\\
\end{eqnarray}
Note that there is a special formal  case when $p=0$ when the turning point
occurs at
\begin{equation}\label{FStp}
\rho^2=\frac{C^2}{q T_{F1}(qT_{F1}-\frac{2C}{L})} \ .
\end{equation}
We see that given turning point is real  when  $C<2\frac{q T_{F1}}{L}$.
We also see from (\ref{FStp}) that  the string can reach
$\rho\rightarrow\infty$ when
\begin{equation}
C_{cr}= \frac{q}{2} T_{F1}L \ .
\end{equation}
In fact, it is easy to see that for $C>C_{cr}$, the expression under
the square root in (\ref{dottfs})
 is always positive for all $\rho$. Hence, for $C>C_{cr}$ and for $p$  the given
configuration can always reach $\rho\rightarrow\infty$.

 Finally, we would like to compare the given result with the analysis
of the motion of fundamental strings in the given background. Recall
that the dynamics of the classical string is governed by the Nambu-Gotto
action
\begin{equation}
S_{NG}=-T_{F1}\int d\tau  \int_0^l d\sigma [ \sqrt{-\det
g_{\alpha\beta}}
+\frac{1}{2}\epsilon^{\alpha\beta}B_{MN}\partial_\alpha
x^M\partial_\beta x^N] \ ,
\end{equation}
where $\epsilon^{\tau\sigma}=-\epsilon^{\sigma\tau}=1$. Now the
equation of motion of $x^M$ takes the form
\begin{eqnarray}
\partial_\alpha[ g_{MN}\partial_\beta x^N
g^{\beta\alpha}\sqrt{-\det
g_{\alpha\beta}}+\epsilon^{\alpha\beta}B_{MN}\partial_\beta x^N]=0 \ .
\nonumber \\
\end{eqnarray}
Consider the time-dependent
ansatz as in case of $D1$-brane
\begin{equation}
x^0=\tau \ , \chi=\sigma \ , g_{\sigma\sigma}=g_{\chi\chi} \ ,
g_{\tau\tau}=g_{tt}+g_{\rho\rho}(\dot{\rho})^2 \ .
\end{equation}
Then the equation of motion for $\chi$ is obeyed automatically while
that for $t$ implies
\begin{eqnarray}
\dot{\rho}^2=(1+\frac{\rho^2}{L^2})
(1-\frac{(1+\frac{\rho^2}{L^2})\rho^2}{(C+\frac{\rho^2}{L^2})^2})
\nonumber \\
\end{eqnarray}
Now the expression on the right has turning point at
\begin{eqnarray}
\rho^2=\frac{C^2}{1-2\frac{C}{L}}  \Rightarrow C<\frac{L}{2} \nonumber \\
\end{eqnarray}
Let us try to integrate the equation of motion for $C=\frac{L}{2}$
when we obtain
\begin{equation}
d\rho\frac{(1+2\frac{\rho^2}{L^2})}{\sqrt{1+\frac{\rho^2}{L^2}}}=dt
\end{equation}
Integrating both sides we obtain
\begin{equation}
\frac{\rho^2}{L^2}=\frac{-1+\sqrt{1+(t-t_0)^2}}{2} \
\end{equation}
and we see that given string reaches $\rho=\infty$ in the limit
$t\rightarrow \infty$. It is also easy to see that for $C>L/2$ there
is no turning point and fundamental string always reaches $\rho\rightarrow\infty$
which is well known  fact \cite{Bachas:2000fr}.

%
%

\end{appendix}

\end{document}